\documentclass[twocolumn]{aastex63}

\usepackage{amsmath}
\usepackage{epsfig}
\usepackage{amssymb}
\usepackage{graphicx}
\usepackage{color}
\usepackage{natbib}

\def\gtaprx {\lower .1ex\hbox{\rlap{\raise .6ex\hbox{\hskip .3ex
	{\ifmmode{\scriptscriptstyle >}\else
		{$\scriptscriptstyle >$}\fi}}}
	\kern -.4ex{\ifmmode{\scriptscriptstyle \sim}\else
		{$\scriptscriptstyle\sim$}\fi}}}
\def\ltaprx {\lower .1ex\hbox{\rlap{\raise .6ex\hbox{\hskip .3ex
	{\ifmmode{\scriptscriptstyle <}\else
		{$\scriptscriptstyle <$}\fi}}}
	\kern -.4ex{\ifmmode{\scriptscriptstyle \sim}\else
		{$\scriptscriptstyle\sim$}\fi}}}

\newcommand{\cutt}[1]{\textcolor{blue}{}}

\newcommand{\Ms}{{\ensuremath{M_{\odot} }}}

\newcommand{\Zs}{\ensuremath{Z_\odot}}

\def\sfr{ \Ms /{\rm yr}\ {\rm Mpc}^{-3} }
\def\inta{\int_{M_{\rm Min}}^{M_{\rm Max}} \Psi (M) dM}
\def\intb{\int_{15\Ms}^{500\Ms} M \Psi (M) dM}

\newcommand{\C}{{\ensuremath{^{12}\mathrm{C}}}}
\newcommand{\Ni}{{\ensuremath{^{56}\mathrm{Ni}}}}

\newcommand{\N}{{\ensuremath{^{14} \mathrm{N}}} }

\begin{document}

\title{Detecting Pair-Instability Supernovae at $z \lesssim$ 5 with the {\em James Webb Space Telescope}}

\correspondingauthor{Enik\H o Reg\H os}
\email{enikoe.regoes@gmail.com}

\author{Enik\H o Reg\H os}
\affiliation{Konkoly Observatory,CSFK, Konkoly Thege M. ut 15-17, Budapest, 1121, Hungary}

\author{J\'ozsef Vink\'o}
\affiliation{Konkoly Observatory,CSFK, Konkoly Thege M. ut 15-17, Budapest, 1121, Hungary}
\affiliation{Department of Optics and Quantum Electronics, University of Szeged, Dom ter 9, Szeged, 6720, Hungary}

\author{Bodo L. Ziegler}
\affiliation{Department of Astrophysics, University of Vienna, Tuerkenschanzstrasse 17, 1180, Vienna, Austria}


\begin{abstract}

Pair-instability supernovae (PISNe) are the ultimate cosmic lighthouses, capable of being observed at $z \gtrsim$ 25 and revealing the properties of primordial stars at cosmic dawn. But it is now understood that the spectra and light curves of these events evolved with redshift as the universe became polluted with heavy elements because chemically enriched stars in this mass range typically lose most of their hydrogen envelopes and explode as bare helium cores.  The light curves of such transients can be considerably dimmer in the near infrared (NIR) today than those of primordial PISNe of equal energy and progenitor mass.  Here, we calculate detection rates for PISNe whose progenitors lost their outer layers to either line-driven winds or rotation at $z \lesssim$ 10, their detection limit in redshift for the {\em James Webb Space Telescope} ({\em JWST}).  We find that {\em JWST} may be able to detect only Pop II (metal-poor) PISNe over the redshift range of $z \ltaprx 4$, but not their Pop III (metal-free) counterparts. 
 
\end{abstract}



\section{Introduction}

Stars with masses of 90 - 260 \Ms\ at the end of their lives are expected to encounter the pair instability (PI), in which temperatures and densities in their cores favor the production of positron-electron pairs \citep{rs67,brk67}.  Pair production comes at the expense of thermal photons and thus pressure support in the core, causing it to contract and its temperatures to rise, which in turn produces more rapid burning, positron-electron pairs and core contraction.  Eventually this process triggers explosive burning in the O and Si layers that is capable of completely destroying the star in a highly energetic thermonuclear supernova \citep[SN;][]{hw02,jw11,chen14c,chen14a}.  

With energies of up to 100 times those of Type Ia SNe, Population III (or Pop III) PISNe can be detected at $z \gtrsim$ 25, revealing the properties of the first stars in the universe \citep{wet12a,wet13c,wet12b,wet13d}.  In contrast, {\em JWST} will not detect core-collapse (CC) SNe beyond $z \sim$ 10 - 15 \citep{wet12c}.  Although Type IIn SNe and gamma-ray bursts (GRBs) could be observed at $z \sim$ 20 - 25, there may be far fewer of them than PISNe at these epochs, depending on the Pop III initial mass function \citep[IMF;][]{wet12e,mes13a}.

But the spectra and light curves of PISNe can change significantly with the metallicity of the progenitor star as the universe gradually became polluted by SNe over cosmic times.  Higher metallicities lead to larger mass-loss rates due to line-driven winds that can strip much of the H layer from the star by the time it encounters the PI (\citealt{lang07,yoon12}; rotation can have a similar effect -- \citealt{cw12,cw12a,cwc13,cw14a}).  The star is therefore much more compact at death, with explosion light curves that are quite different from those of Pop III PISNe.  Simulations show spectra that high-$z$ PISNe whose progenitors have lost at least part of their outer layers are much dimmer in the NIR today than Pop III PISNe of equal energy and mass because the fireball cools to temperatures at which its emission peaks in the NIR at much smaller radii \citep{wet13e,smidt13a,smidt14a}.  Consequently, it can be more difficult to observe PISNe at $z \lesssim$ 10 than at $z \sim$ 20.

What then are the prospects for detecting PISNe at these lower redshifts in the coming decade with {\em JWST} and ground-based extremely large telescopes (ELTs)?  Although they are dimmer, there may be more of them because of the rise in cosmic star formation rates (SFRs) at $z <$ 10 inferred from the luminosity functions of early galaxies \citep{camp11}, from GRB rates \citep{idf11,re12}, and from simulations \citep[e.g.,][]{wise12,jdk12,pmb12,xu13,haseg13,mura13,magg16,xu16}.  Previous estimates of PISN detection rates out to $z \sim$ 20 \citep{sc05,wl05,wa05,pan12a,hum12,tet12,ds13,ds14,hart18a,ryd18a,moriya19} may be too high at $z <$ 10 because they all used light curves of Pop III explosions, which are much brighter than the more realistic spectra of explosions of partially stripped stars. 

Although only a few PISN candidates have been found at low redshifts to date \citep{gy09,cooke12,ogle} the unprecedented sensitivities of {\em JWST} and the ELTs could reveal far more of them, probing SFRs at higher redshifts than ever before and the stellar populations of the first galaxies.  Here, we calculate detection rates for PISNe at $z \lesssim$ 10 for {\em JWST} for compact progenitors that have lost some of their envelopes to either line-driven winds or rotation.  In Sections~2 - 3 we describe our PISN spectral models and how they are cosmologically dimmed, redshifted and convolved with filter functions, and cosmic SFRs in Section 4 to obtain detection rates.  In Section~5 we discuss these rates and conclude in Section~6.
Some of our light curves are in the Appendix.

\section{Numerical Method}

Our source frame spectra for $z \lesssim$ 10 PISNe (essentially, bare He cores) were calculated in three stages.  First, the progenitor star was evolved from the zero-age main sequence to the onset of the PI, explosion, and the end of all thermonuclear burning in the GENEVA and KEPLER codes \citep{e08,kep2} and in the MESA and FLASH codes \citep{paxt11,paxt13,Fryx00,dub09}.  Each SN was then evolved from shock breakout from the surface of the star to expansion into the IGM with the Los Alamos radiation hydrodynamics code RAGE \citep{rage}.  All explosions were evolved for 3 yr after breakout. RAGE profiles for the PI SNe were then post processed with the Los Alamos SPECTRUM code to obtain spectra for the fireball in the rest frame \citep{fet12}. 

We consider two sets of progenitors: 0.14 - 0.43 \Zs\ 150 - 500 \Ms\ stars that have lost their hydrogen layers to strong winds \citep{wet13e} and 90 - 140 \Ms\ Pop III stars that have lost their outer envelopes to rotation \citep{smidt14a}.  We adopt the spectra for the explosions in low densities in \citet{wet13e} to obtain upper limits to their luminosities in the NIR today.  The properties of both sets of PISNe (progenitor mass, metallicity, explosion energy and \Ni\ yields) are summarized in Tables~1 of their respective papers, where we also lay out in detail our stellar evolution and burn models, radiation hydrodynamical simulations, and spectrum calculations \citep[see][for other PI SN light curves for a variety of progenitor metallicities and rotation rates]{kasen11,det12,kz14a,kz14b,jer16,gil17,kz17}.

Metals lead to mass loss which reduces the diffusion timescale. A dense circumstellar envelope increases diffusion timescales by contributing to mass. 
Pop III stars explode in low-density relic H II regions while PISNe in the local Universe may occur in dense winds and bubbles blown by the star. 
Collision with prior outburst also adds to the luminosity. 

Most important for estimating metallicity thresholds for PISNe are the mass loss rates of very massive stars during their evolution.
\citep{georgy17} find that H-rich PISN could occur at metallicities as high as \Zs /3 (lower for rotating).
Magnetic field could suppress mass loss even for higher (solar) metallicity.
About 7\% of O-stars have important surface dipolar magnetic field at solar metallicity. Such surface magnetic fields can quench the mass loss from the star \citep{petit17}.
This mechanism can form PISNe at higher metallicity than usually thought, and could explain SLSNe in the local Universe.

The 150 and 200 \Ms Pop II models have metallicity Z = 0.14 \Zs corresponding to the Small Magellanic Cloud. The 500 \Ms model has metallicity Z = 0.43 \Zs corresponding to the Large Magellanic Cloud.
 
\section{JWST magnitudes and colors}\label{sec:obs}

\begin{figure*}
    \centering
    \includegraphics[width=14cm]{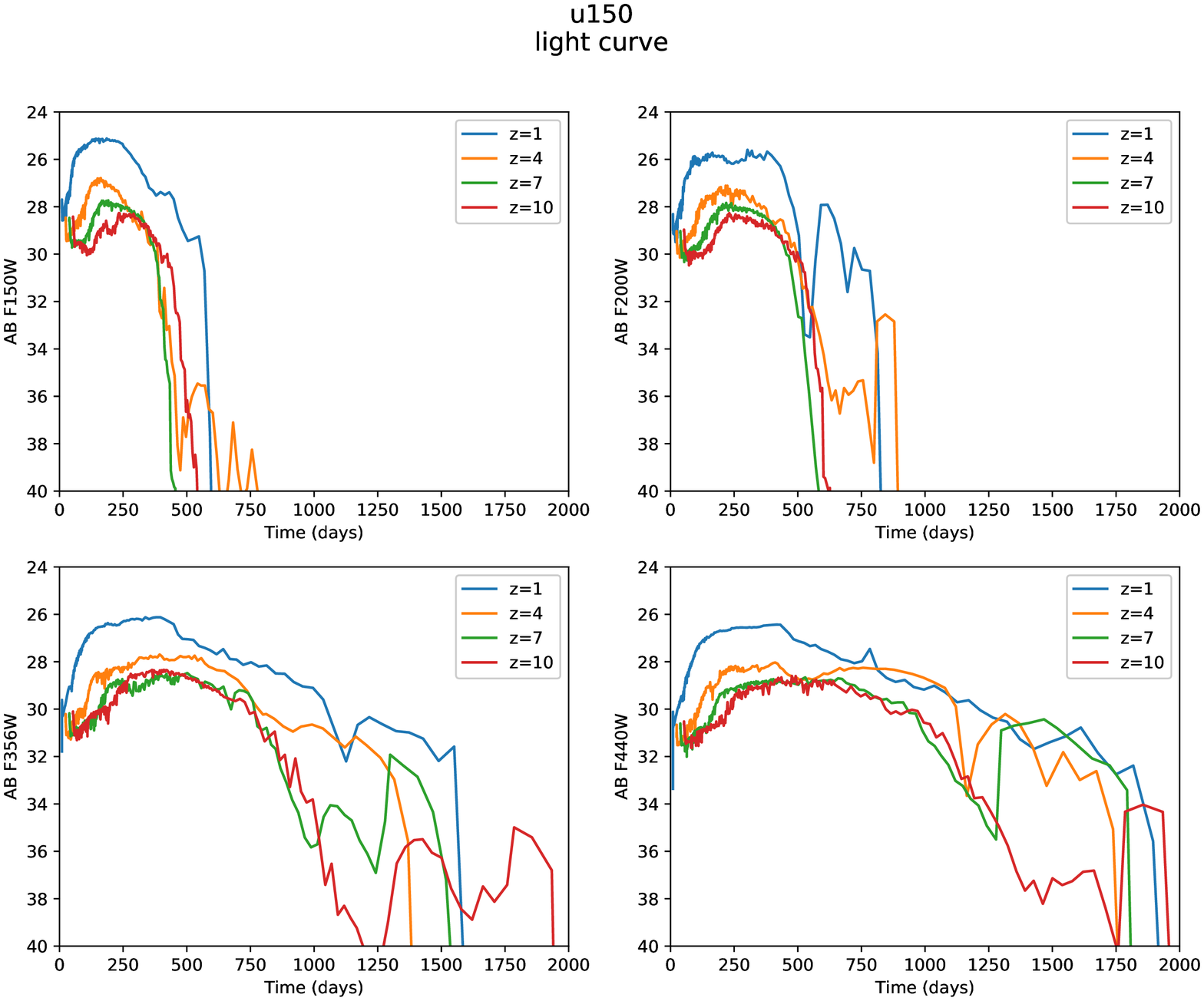}
    \includegraphics[width=14cm]{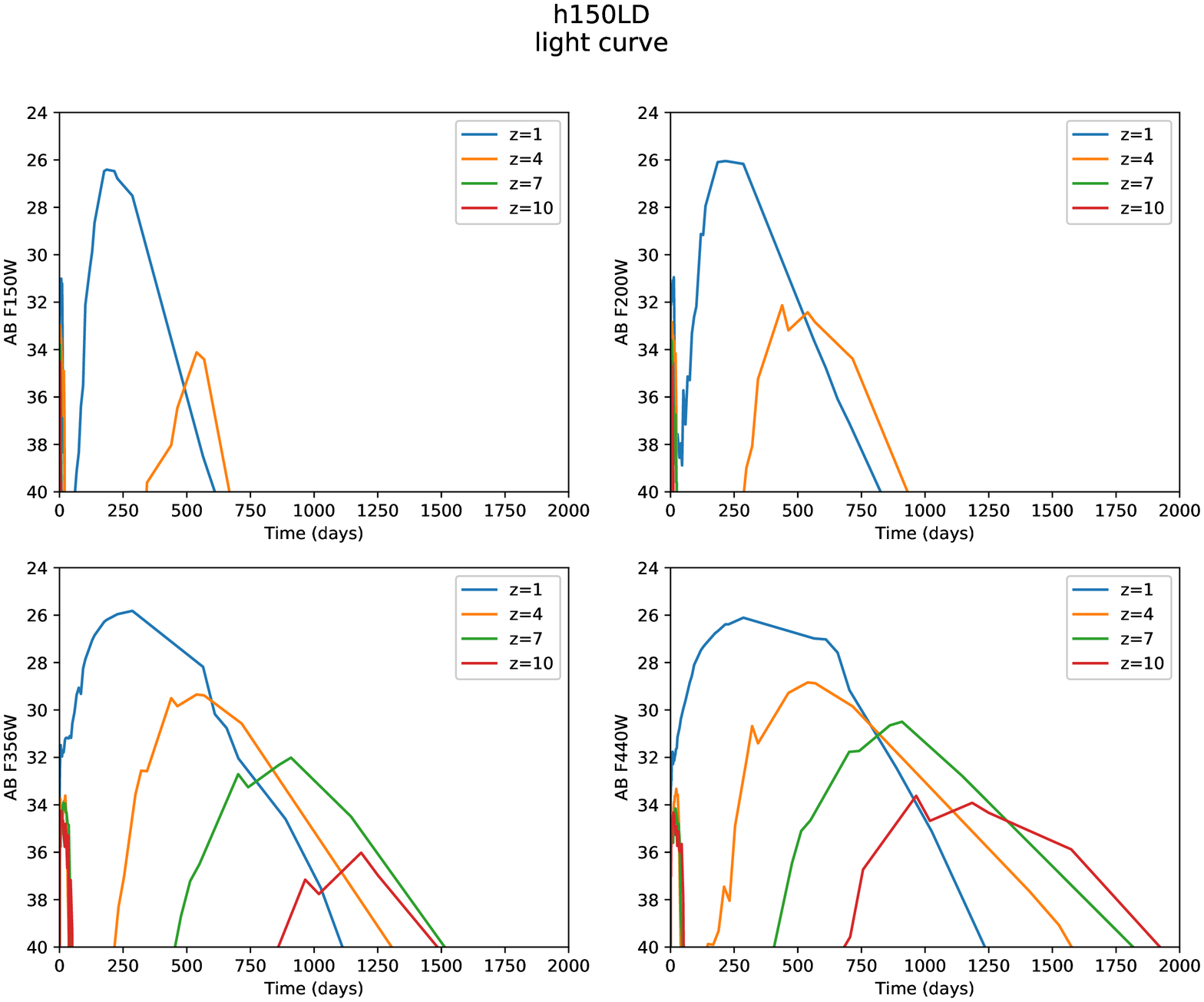}
    \caption{Light curves of 150 \Ms PISN models at various redshifts and {\em JWST} filter wavelengths. Upper panel: 
    Population III (zero metallicity) model; lower panel: Population II, Z = 0.14 \Zs model.}
    \label{fig:popiii-lc1}
\end{figure*}

\begin{figure*}
    \centering
    \includegraphics[width=14cm]{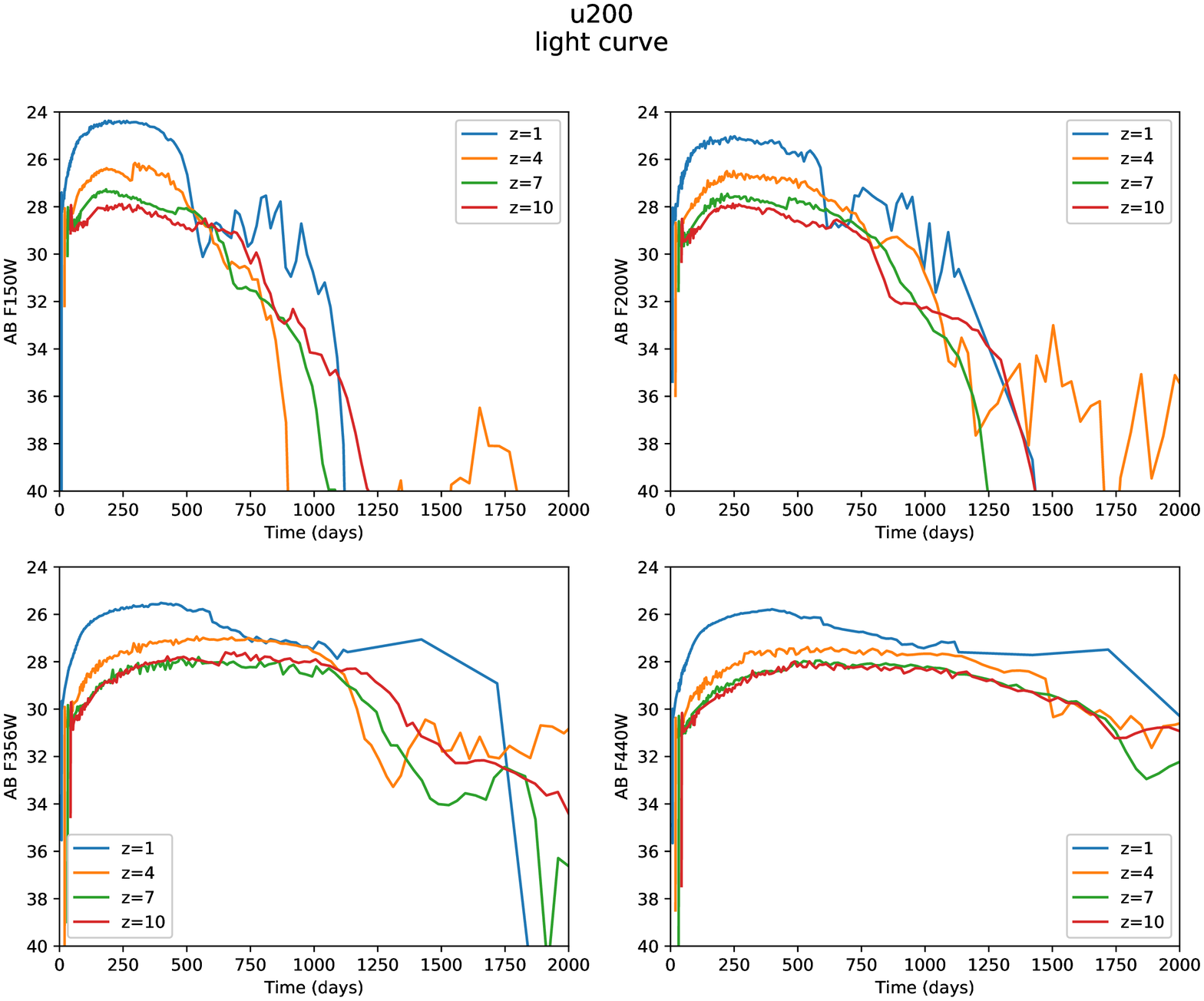}
    \includegraphics[width=14cm]{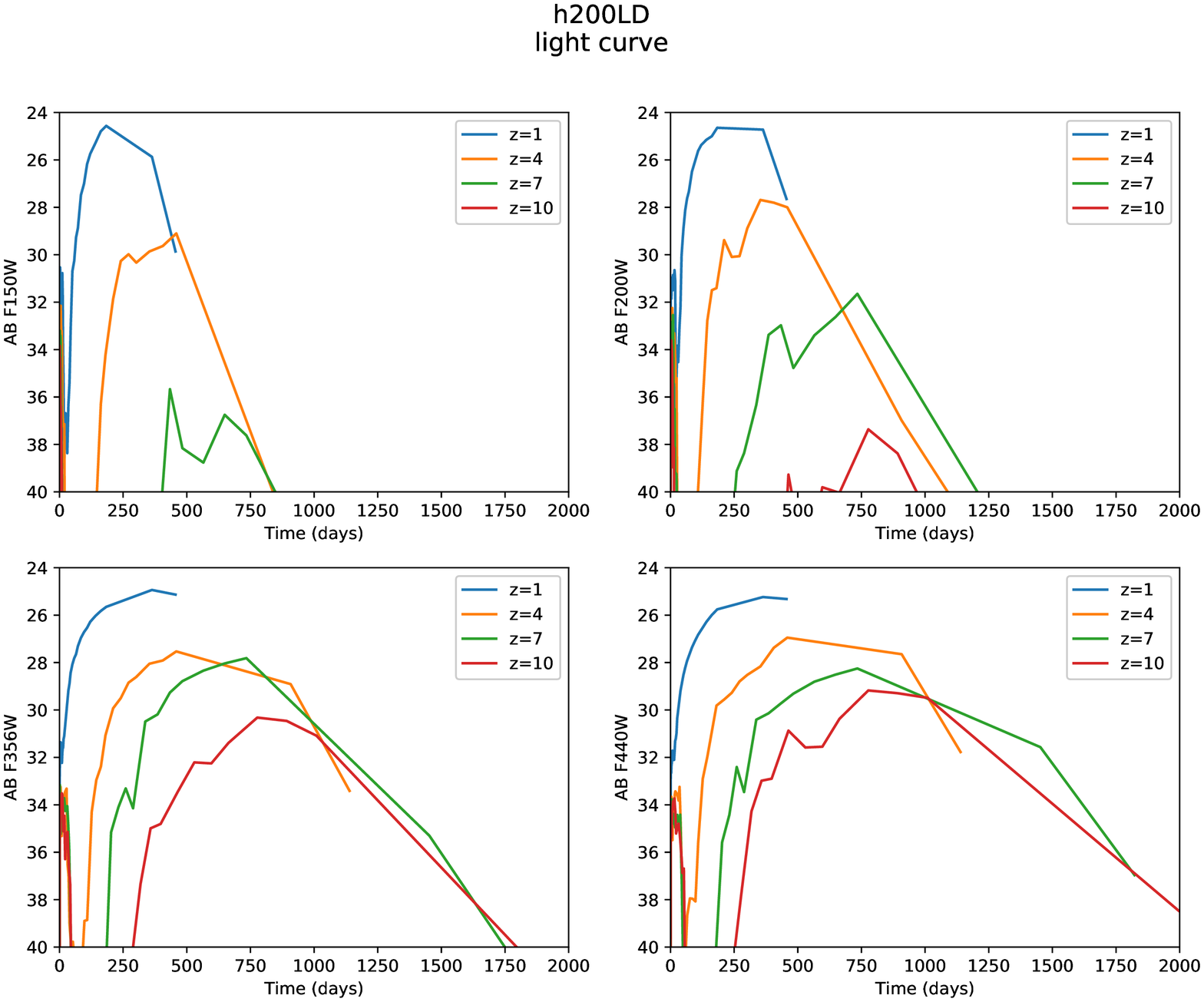}
    \caption{Same as Figure~1 but for 200 \Ms models.}
    \label{fig:popiii-lc2}
\end{figure*}

\begin{figure}
    \centering
    \includegraphics[width=8.5cm]{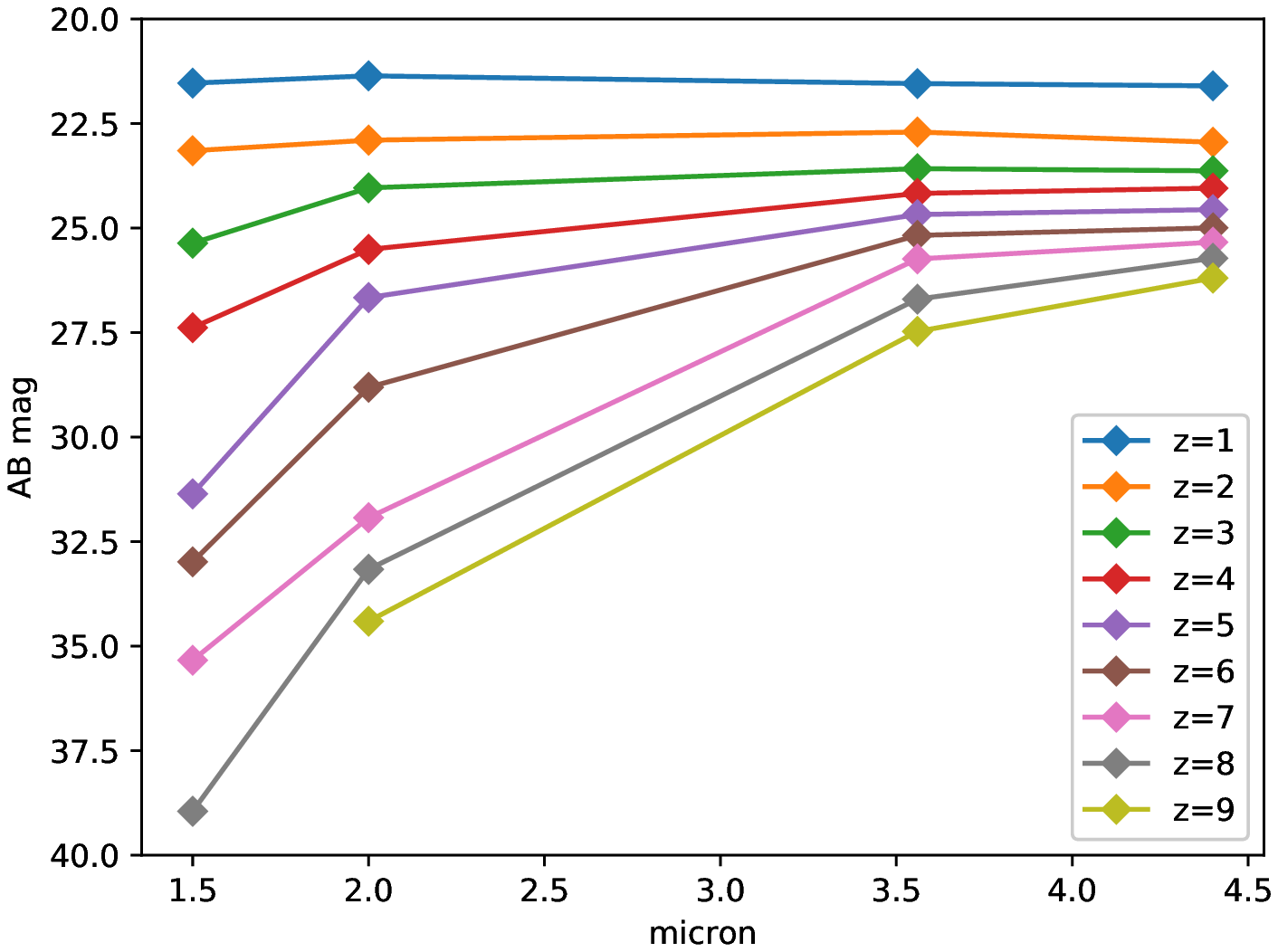}
    \caption{Peak magnitudes of Population II, Z = 0.07 \Zs PISN models of 250 \Ms by  \citet{ch19} at various redshifts as function of NIRCam filter wavelength.}
    \label{fig:popii-lc2}
\end{figure}

\begin{figure}
    \centering
    \includegraphics[width=8.5cm]{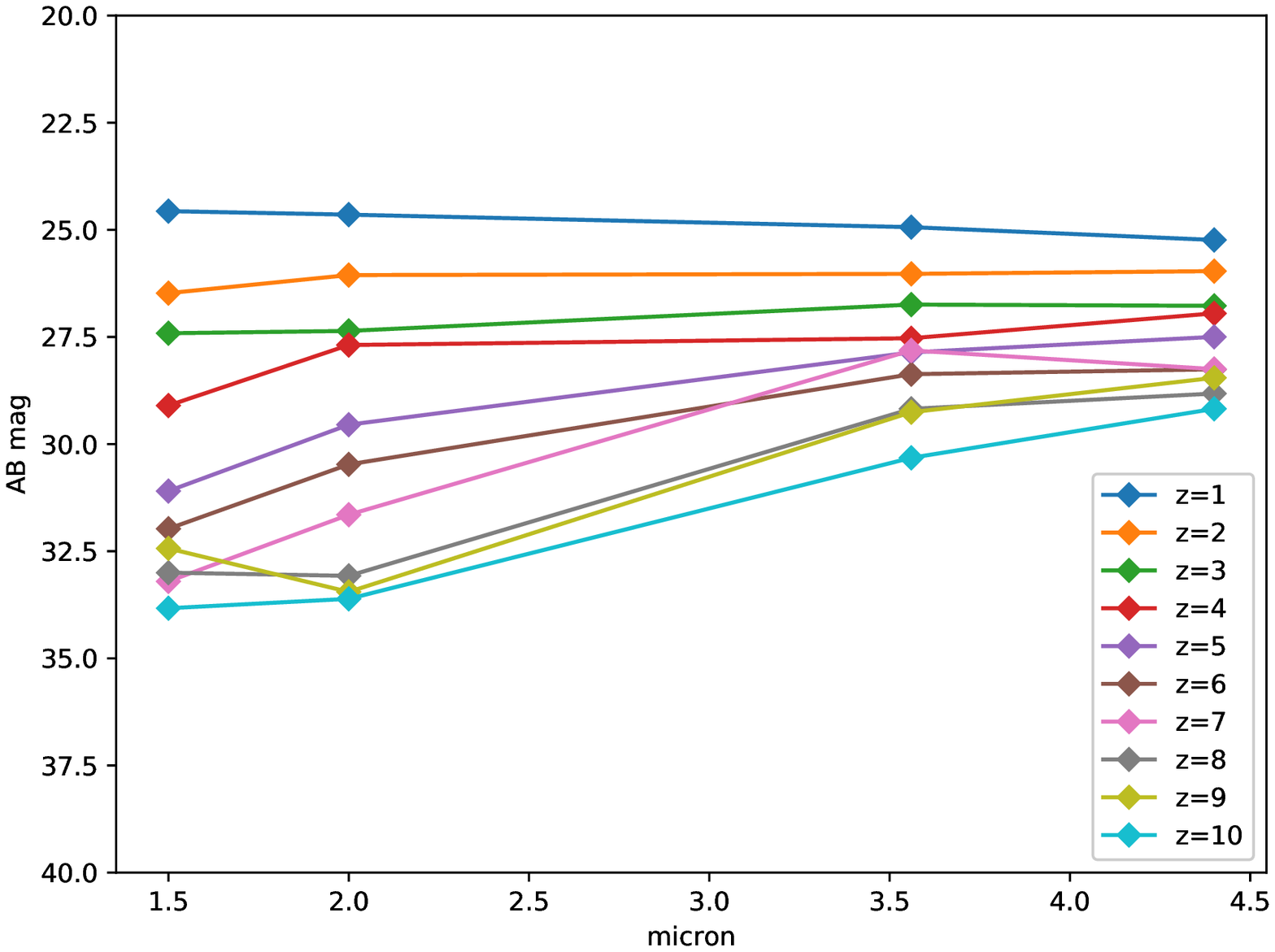}
    \caption{Peak magnitudes of Population II, Z = 0.14 \Zs PISN models of 200 \Ms \citep{wet13e} at various redshifts as function of NIRCam filter wavelength.}
    \label{fig:popii-lc3}
\end{figure}

\subsection{Observational constraints}\label{sec:obs2}

In this paper we consider the observational conditions of the First Lights at REionization (FLARE) project as proposed by \citet{flare}. FLARE intends to discover and classify luminous transients above $z > 2$ redshifts utilizing repeated deep observations with {\em JWST} NIRcam within the {\em JWST} Continuous Viewing Zone near the North Ecliptic Pole (NEP). The total area of the 3 year-long survey is $\sim 300$ arcmin$^2$, which will be monitored in 4 NIRCam filters (F150W, F200W, F356W, F440W) with a $\sim 3$ month cadence. With the proposed conditions the limiting magnitude for a $\sim 3 \sigma$ detection will be $\sim 27$ AB-mag. See \citet{flare} and \citet{rv} for more details.

\subsection{Light curves}\label{sec:lc}

In Figure 1 (upper panel), Figure 2 (upper panel) (and Figures 11-12 in the Appendix) we show synthetic light curves of zero metallicity Pop III PISN models in the F150W, F200W, F356W and F440W NIRCam filter bandpasses for 150 M$_\odot$, 200 M$_\odot$ (and 175 \Ms, 225 M$_\odot$, 250 M$_\odot$) progenitors, respectively. Different redshifts up to $z = 10$ are color-coded, and the time since explosion is expressed in observer's frame days. These light curves were calculated by convolving the time-dependent spectra of these model SNe with the bandpass functions of the NIRCam filters, after redshifting and scaling them by the luminosity distance. We adopted the standard $\Lambda$-CDM cosmology with the following parameters:  \(\Omega_m=0.315, \Omega_\Lambda=0.685, H_0=67.4\) (Planck collaboration, 2018).

As seen in these Figures, Pop III PISNe just barely reach the $\sim 27$ mag detectability limit for $z < 4$ events, and escape detection in all JWST/NIRCam filters at higher redshifts. Usually PISNe are expected to occur at redshifts higher than this limit, i.e. at $z > 6$. However, if there were some Pop III stars formed at lower redshift (i.e. at later epochs on the cosmic timescale) than the majority of them, {\em JWST} might detect a few of those. 

In addition to Pop III models, massive stars having Pop II chemical composition may also produce PISNe. Pop II and Pop III models may have quite different colors and spectra due to primarily the presence of metals in the ejecta of the Pop II events.  In addition, Pop II PISNe do not suffer from strong Lyman~$\alpha$ absorption in the intergalactic medium, although this depends on redshift, i.e. on the ionisation degree of the intergalactic medium. 

Figure 1 (lower panel) and Figure 2 (lower panel) exhibit light curves of Pop II (Z = 0.14 Z$_\odot$ metallicity, low density envelope) models assuming 150 and 200 M$_\odot$ progenitor masses, respectively. 

Figure 1 is a direct comparison of PISN light curves from Pop II and Pop III (for 150 \Ms ). While the upper panel is metal free,
the lower panel shows PISNe from metal enriched stars for the detection with JWST.
Figure 2 is the same comparison of Pop II and Pop III (as Figure 1) for 200 \Ms .

In addition, we also examine the Pop II, Z = 0.07 Z$_\odot$ model of \citet{ch19} having 250 M$_\odot$ progenitor mass. Peak AB-magnitudes from the latter model as a function of filter wavelength, color-coded for different redshifts, are shown in Figure~\ref{fig:popii-lc2}.

Figure 4 displays peak magnitudes of the Pop II, $Z = 0.14 \Zs$ 
model \citep{wet13e} as function of filter wavelength and redshift.

It is seen in Figure~3 that Population II, Z = 0.07 \Zs PISNe of 250 \Ms \citep{ch19} may be detected between $1 < z < 4$ redshifts in the FLARE survey, if we assume $\sim 27$ AB-mag as the $3 \sigma$ detection limit in all filters.  This is different from the case of the zero metallicity Pop III PISNe, because the peak magnitudes of the simulated light curves of the latter seem to be somewhat 
fainter (see also Figures~1 and 2, lower panels).

Nevertheless, both sets of light curves show very slow evolution: they are expected to stay near maximum light for an order of $\sim 100$ days. Time dilation stretches with a factor of $1+z$ in the observer's frame, causing visibility at higher redshifts. This will increase the detection probability for such events, because if such a transient is detected within the survey field, it will likely remain detectable for more than a year in the observer's frame. 

\subsection{Color-color diagram}\label{sec:cc}

Color-color plots are useful tools for diagnosing various types of transients having different SEDs.  They are distance-independent, thus, objects appearing at different distances are comparable. They do, however, preserve the redshift-dependence of the colors, therefore the separation of transients at low- and high redshifts remains possible. 

Color-color diagrams utilizing various {\em JWST} filter combinations have been proposed for classifying transient objects. For example \citet{tet13} suggested the application of F200W-F277W versus
F277W-F356W as a diagnostic tool for discovering SLSNe. More recently, \citet{rv} found that
Type Ia SNe at $z > 1$ redshifts can be likely identified on the F200W-F444W vs F150W-F356W color-color diagram. 

\begin{figure}
    \centering
    \includegraphics[width=8.5cm]{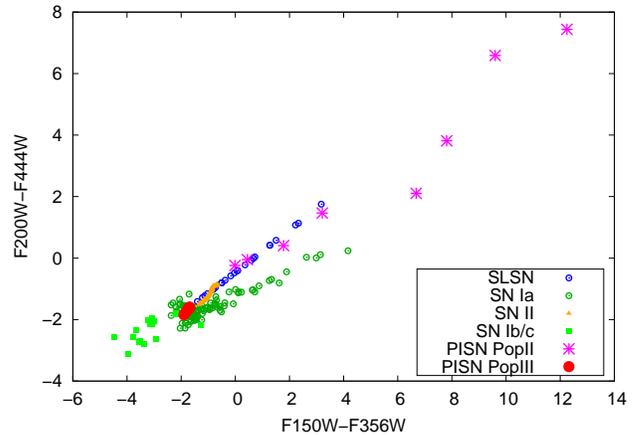}
    \caption{{\em JWST} color - color diagram of Pop II (magenta) and Pop III (red symbols) PISNe and
    various other types of SNe around maximum light. The position of SNe Ia (circles), Ib/c (asterisks), II-P (squares) and SLSNe (triangles) on the {\em JWST} color-color (F200W-F444W vs. F150W-F356W) diagram. Type Ia SNe are shown as circles, while the squares, crosses and asterisks indicate the low-redshift Type II-P and Ib/c SNe that may contaminate the observed sample.}
    \label{fig:color-color}
\end{figure}

The position of PISNe having Pop~III as well as Pop~II metallicity are plotted on the F200W-F444W vs F150W-F356W  color-color diagram in Figure~\ref{fig:color-color}. Pop~II/Pop~III PISNe are represented by magenta/red symbols, respectively. Each symbol corresponds to a 
150 (Pop III) or 250 (Pop II) M$_\odot$ model around maximum light but at different redshifts between $z = 1$ and 10. Also shown are the various types of SNe modeled by \citet{rv}.

It is seen that Pop II PISNe occupy nearly the same range on the color-color plot as the SLSNe, which is not unexpected as both types of transients are thought to be originated from very massive progenitors. Pop III PISNe are clustered in a small region located in the middle of the area occupied by the hydrogen-poor SNe Ia and Ib/c, thus, metal-free PISNe, if they were observable, would look nearly the same as the Type Ia or Ib/c SNe at lower redshifts. This means that the contamination by low-redshift transients will probably be a serious observational challenge, at least as far as the identification of Pop III PISNe are concerned. 

High-redshift observations of time domain astronomy are less demanding on cadence than local ones due to the time dilation, and NEP revisits can be proposed with normal STScI planned annual cycles.
Spectroscopic followup of targets of opportunity with NIRSPEC and ELT
instruments may also be a modest imposition on these facilities.
Deep ground based reference fields at shorter wavelengths should be initiated immediately to eliminate foreground objects (see \citet{flare}). The Subaru Strategic Program \citep{aihara18} is a model for such data and adding the NEP to the current fields is a way. X-ray followup of JWST high-redshift transients will also be vital. 
Spectroscopic followup can help to identify and distinguish PISNe from other sources.

Unambiguous spectral identification at any redshift is unlikely.
A light curve could be more discriminant, if the inferred nickel mass was high.
There has not been any candidate event that sufficiently matches PISN model spectra.
If one got one event to match any model spectra there/that would be a strong case.
A model matching spectrum observed would indicate a PISN.
There are many neutral lines in the spectra: Fe, Si, O; also Ca II, Mg II, S II.
In general lower mass events have more lines (a forest of lines, \citet{jer16}).
Mixing affects spectroscopic and color evolution.
Strong features of intermediate mass elements dominated by silicon, magnesium and oxygen
reach intensities depending on the extent of mixing.

\section{PISN Rates}\label{sec:rates}

\subsection{Rates of Pop II and Pop III events}\label{sec:rates2}

Theoretical models of star formation suggest that
there is a transition from metal-free Pop III to metal-rich Pop I-II star formation. 
This transition occurs at $z > 3-4$ \citep{tfs07, maio10,tss09}.
There remain pristine regions where Pop III star formation and PISN prevail in patches even around redshifts $z \sim 3$.
The metal enrichment is competing against cosmological expansion, so the pockets of metal-free gas remain isolated to some extent. Still, even these pockets are polluting themselves.
{\em JWST} will help to constrain how long Pop III star formation continues. 

The Pop III star formation rate itself depends upon the PISN rate \citep{maio10}.
Their simulations show that it is rather independent from
many parameters, e.g. the critical metallicity (at which metal cooling
dominates over the molecular Lyman-Werner background) for
transition, IMF slope and range, SN/PISN yields or star formation threshold.

\subsection{Effects of the Initial Mass Function IMF}\label{sec:imf}

As massive stars have short life times, which can be approximated as $\sim 30 \times  (M/8M_{\odot})^{-2.5}$ Myr, the PISN rate, $\dot n$, traces the star formation rate (SFR) as a function of cosmic time or redshift.

For a particular mass range and explosion type
\begin{equation}
    \dot n_{\rm SN} (z) = {\rm SFR} (z) \times \frac{\inta}{\intb}
\end{equation}
where $\Psi (M)$ is the IMF \citep{ds14}.

We assume that the IMF is either Salpeter, i.e. $dN/dM \propto M^{-2.35}$ or flat, $dN/dM$ = const.
These choices are motivated by cosmological simulations.

The lower and upper mass limits for PISN are M$_{\rm min}=140\Ms$,
M$_{\rm max}=260\Ms$, or for Pop III may be down to
M$_{\rm min}=85\Ms$ if they are rapidly rotating \citep{cw12}.

\subsection{Effects of the Star Formation Rate}

For Pop II (metal-poor) progenitors we assume that the cosmic star formation rate is similar to the one used in many previous studies, e.g. in \citet{hopbea06}.
The IMF is Salpeter in these Pop II models, and the PISN mass limits are 140 and 260 $M_\odot$ as above.

For Pop III (metal-free) stars the SFR is mostly known from simulations \citep[e.g.][]{ds14} 
with some observational hints.
We assume a flat IMF for such progenitors, and $M_{\rm min}$ can be chosen as 85 $M_\odot$ \citep{cw12}. 

For Population III,
according to \citet{ds11}, \citet{wise12} and \citet{xu13}, by $z\sim 15$ the SFR is $\sim 10^{-4}\sfr$.
\citet{xu13} predict $< 10^{-5}\sfr$ by $z\sim 8$ (and flat from $z\sim 17$).
\citet{bouw12} constrain $\mathrm{SFR} \sim 10^{-2}\sfr$, lower than Pop II but
probably higher than Pop III.
In \citet{yang15} Figure 2 indicates a maximum of $\sim 10^{-2}\sfr$
for the SFR.

Models of Pop III.2 have at least an order of magnitude smaller star formation rates than those of Pop II.  For Pop III.2, that is primordial but irradiated by other stars, \citet{bouw12} displays $\leq 10^{-2}\sfr$ at $z \sim 7$ and $\sim$ $ 2 \times10^{-2}\sfr$ at $z\sim 5$.

Theory predicts $\sim 10^{-3}\sfr$ for Pop III.2 at $z\sim 5$ ($\sim 40$ Pop II)
and $\sim 10^{-2.5}\sfr$ by $z\sim 7$ (a larger Pop II ratio).

Figure 6 is an overview of the various star formation rates as a function of redshift.
The transition from Pop III to Pop II star formation (the redshift above which Pop III star formation rate dominates) is highly uncertain
and may be in the range 10 - 15.

\begin{figure}
    \centering
    \includegraphics[width=8cm]{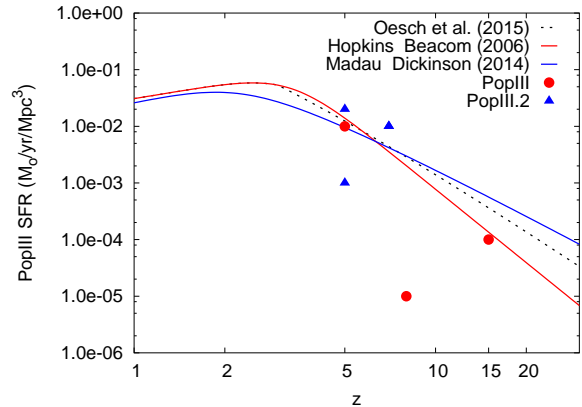}
    \caption{Overview of the various star formation rate densities as function of redshift (Pop III, Pop III.2, Pop II). 
    The red markers at $z > 5$ are from \citet{xu13}, the lower blue triangle at $z = 5$ from \citet{yang15}. The rest of the symbols are \citet{bouw12}}
    \label{fig:popiii-sfr}
\end{figure}

\subsection{Normalisation}\label{sec:norm}

\begin{figure*}
    \centering
    \includegraphics[width=8cm]{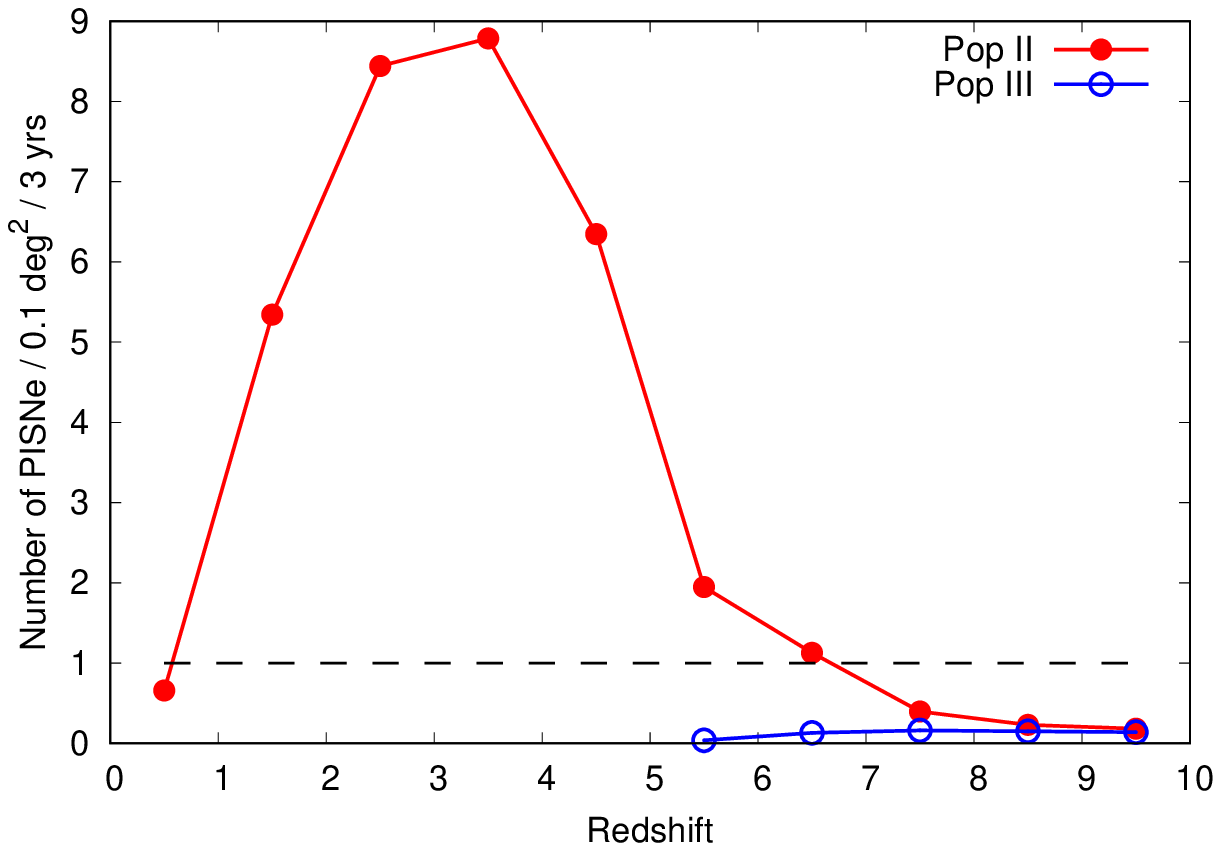}
    \includegraphics[width=8cm]{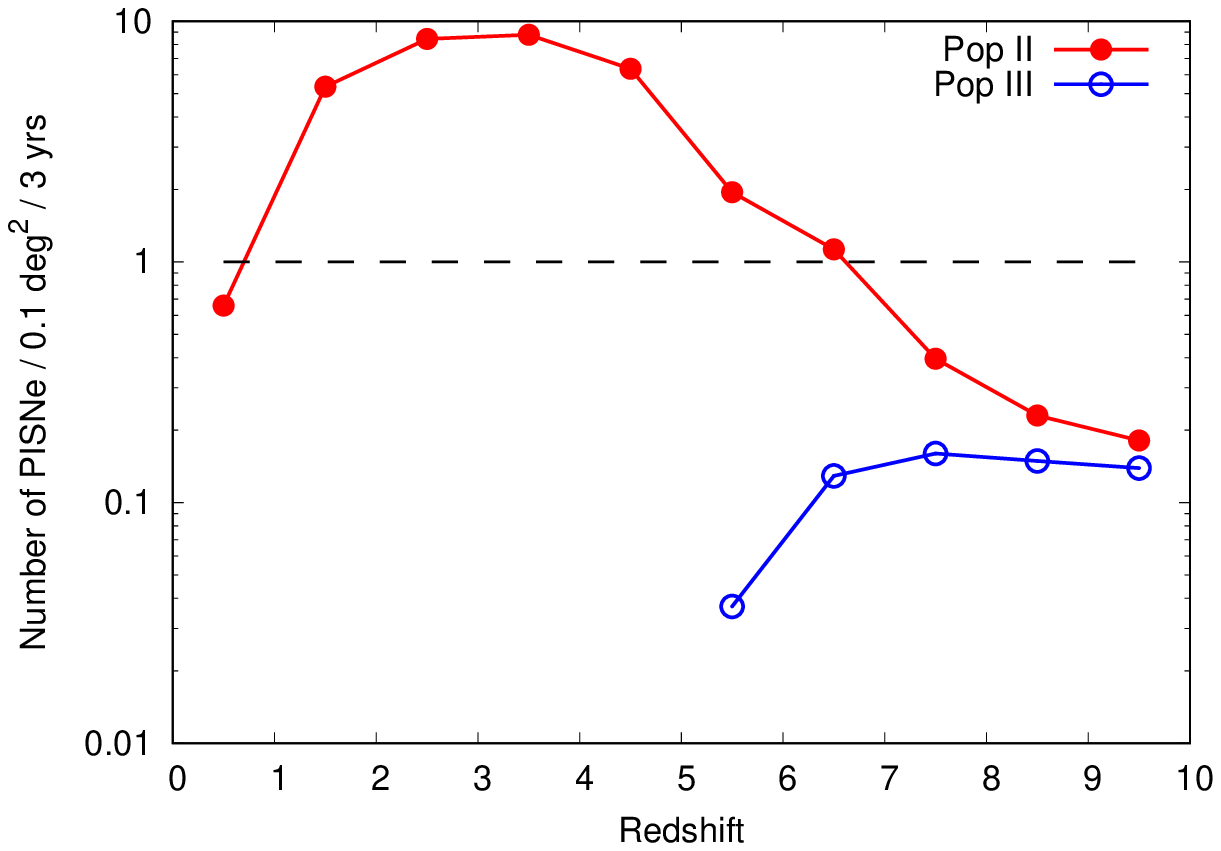}
    \caption{Expected numbers of Population II (red dots/curve) and Pop III PISNe (blue circles/curve) in the FLARE survey. The right panel shows the same data on logarithmic vertical scale. The dashed horizontal line marks the limit of one new event during the survey time.}
    \label{fig:pisn-counts}
\end{figure*}

The Population III PISN rates were calculated with a semi-analytical model that utilises a
cosmologically representative set of halo merger trees with detailed
prescriptions for radiative and chemical feedback on star formation in the
halos \citep{rz10, til15a, magg16}. This model assumes flat IMF.

While the integration over the mass function in Eqn. 1 is not very sensitive to the normalisation by the IMF lower mass limit in the denominator for flat IMF for Pop III stars, for the Salpeter IMF for Population II it is. \citet{wl05} integrate the IMF between 0 and infinity and consider bimodality, but see other constraints on the range and lower limit as well.

To avoid difficulties, we normalise our Pop II PISN rates from
extrapolating observations of GRBs, which are used as tracers of
early star formation \citep{re12}.

The expected number of PISNe between redshifts $z$ and $z + \Delta z$ can be calculated by integrating their rate to get
\begin{equation}
N_{\rm PISN} = \Omega T \int_z^{z+\Delta z} \frac{\dot n_{\rm PISN}(z)}{1+z}\,\frac{\mathrm{d}V}{\mathrm{d}z}\,\mathrm{d}z,
\end{equation}
where $\Omega$ is the survey area, $T$ is the survey time and 
$\mathrm{d}V/\mathrm{d}z$ is the differential comoving volume \citep[e.g.][]{rv}. Note that this gives the number of new events within the survey area during the survey time, but does not take into account cadence or any other observational details that may affect the detection efficiency. With all these other factors included, the number of actual detections will be lower. We will consider the effect of cadence and sensitivity in Section~\ref{sec:simul}.

\subsection{The expected number of PISNe}

The expected number of PISNe within the FLARE survey field during the survey time are plotted in Figure~\ref{fig:pisn-counts}, where the red dotted curve shows the expected number of Pop II PISNe as a function of redshift, while the blue curve is the same but for Pop III PISNe. 

It is seen that a few Pop II PISNe can be expected to occur per unit redshift even within the small ($\sim 300$ arcmin$^2$) FLARE survey field during the 3 year-long survey. This is not the case for Pop III PISNe, however, as their expected numbers are below 1 in the whole redshift range.

Recently, similar numbers were found by \citet{rv} for the expected counts of Superluminous Supernovae (SLSNe) during the FLARE survey. Even having the preference of the metal-poor environment of SLSNe taken into account, they were able to predict only a few SLSNe in the FLARE FoV during the 3 years survey time. Considering that the expected progenitors of PISNe and SLSNe are thought to be very massive stars with a similar mass range, our result above for PISNe is in accord with the prediction of \citet{rv}. 

Integrating the Pop II PISN rate (the red curve in Figure~\ref{fig:pisn-counts}) over redshift gives the total number of such PISNe as $N \sim 30$ within the survey field during 3 years. We use this number to estimate the number of actually detected PISNe during the survey in the next section.

\subsection{Simulation}\label{sec:simul}

\begin{figure*}
    \centering
    \includegraphics[width=12cm]{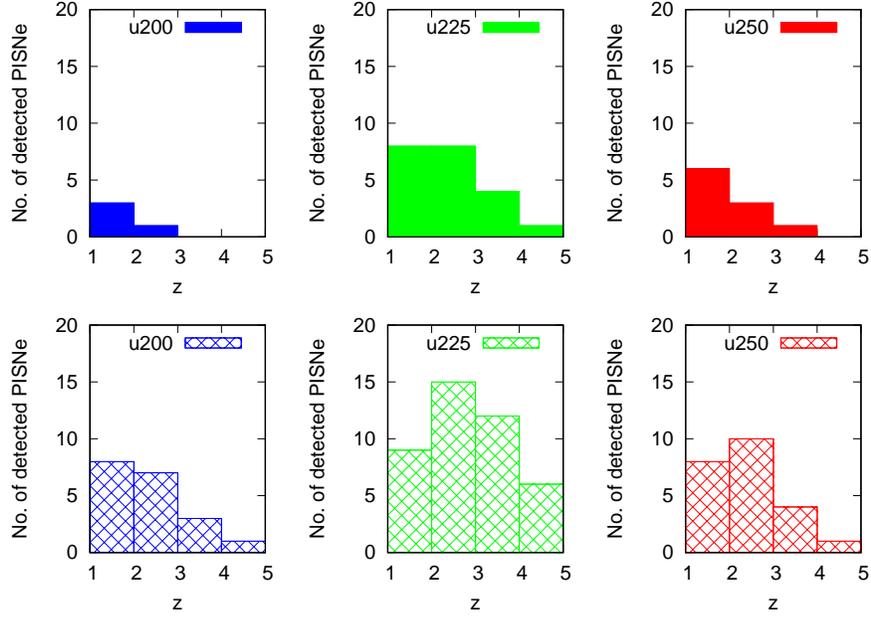}
    \caption{The number of detected PISNe in the FLARE survey simulation, for the zero metallicity 200, 225 and 250~M$_\odot$ models, as a function of redshift. The upper panels corresponds to the ``strong detection'' criterion of \citet{rv}, while the lower panels show the numbers in the``weak detection'' case. }
    \label{fig:pisn-simul-strongweak}
\end{figure*}

\begin{figure*}
    \centering
    \includegraphics[width=16cm]{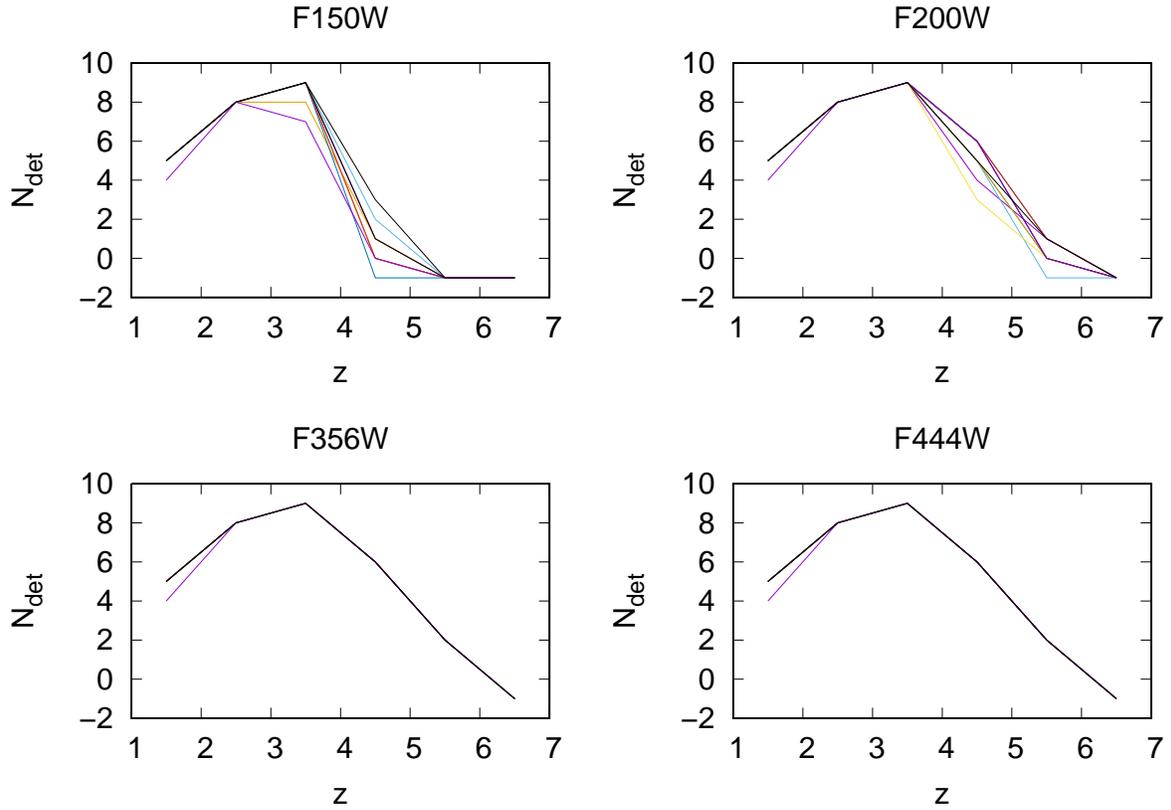}
    \caption{The number of ``weak'' detections in the FLARE survey simulation, using the 250~M$_\odot$ model of \citet{ch19}, as a function of redshift in different filters. Different colors represent 10 random realizations of the simulation.}
    \label{fig:pisn-simul-distr}
\end{figure*}

\begin{figure}
    \centering
    \includegraphics[width=8cm]{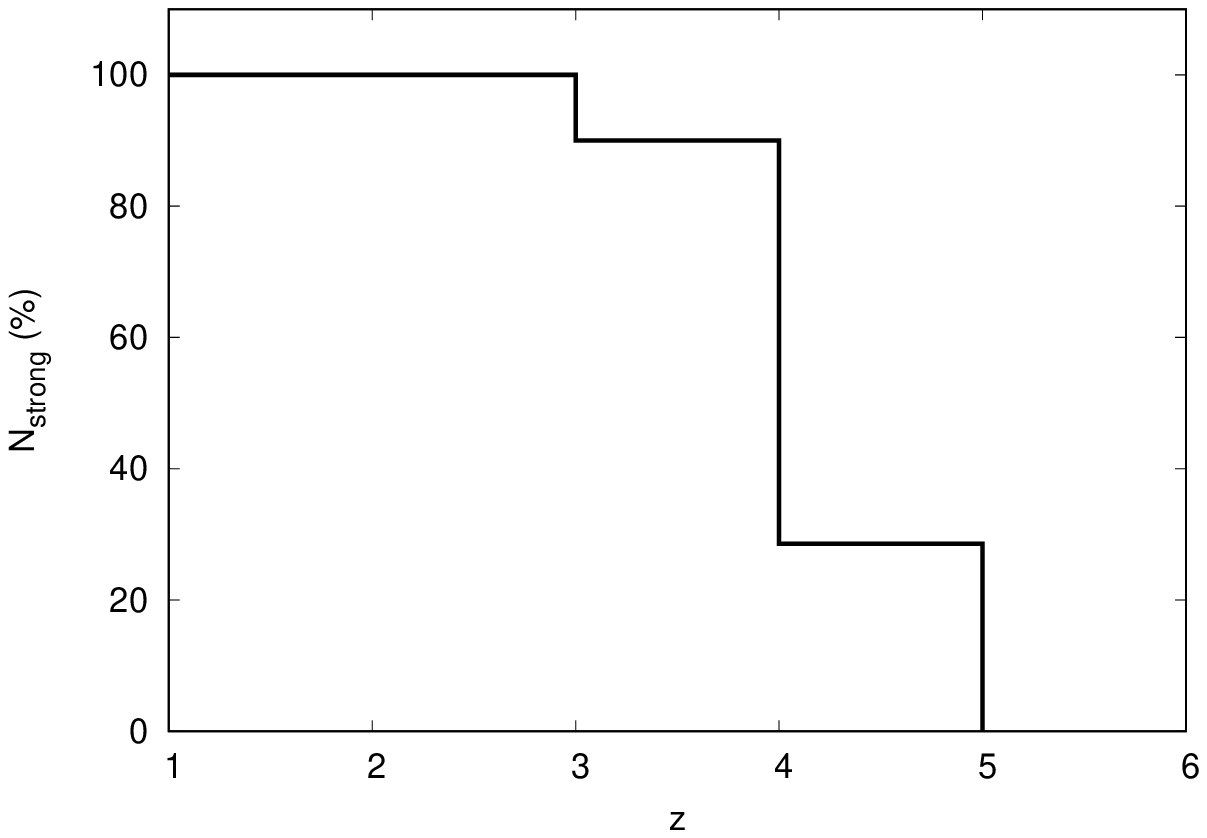}
    \caption{The percentage of strong detections in the FLARE survey simulation as a function of redshift.}
    \label{fig:pisn-simul-strong}
\end{figure}

Following \citet{rv}, we simulate the FLARE survey observations with {\em JWST} by utilizing the {\tt sncosmo}\footnote{https://sncosmo.readthedocs.io/en/v2.0.x/} code \citep{sncosmo16}. We draw random samples of $2 \times 30 ~=~ 60$ PISNe with explosion times distributed uniformly between [$-1000$:1000] days with respect to the starting date of the survey, $t_{start} = 0$ day (the simulation is started $1000$ days before $t_{start}$ in order to take into account those events that are already present in the observed field after the initiation of the survey).

We use 90 days cadence during the 3 year-long survey in the observer's frame, and sample the light curves from the time-evolving SEDs of the various PISN models with the four selected NIRCam filters (see Section~\ref{sec:obs}) using {\tt sncosmo}. We assume that the redshift distribution follows the one for the Pop II events as shown in Figure~\ref{fig:pisn-counts}.

First, we consider the zero metallicity models (cf. Section~\ref{sec:obs}) having 200, 225 and 250 M$_\odot$ masses.
Because the $z>6$ events are not expected to reach the detection limit of 27.3 AB-magnitude of the FLARE survey, we place them at closer distances and assume the same redshift distribution for them as that of the Pop II events. For simplicity, in a random realization of the simulation we use the same mass for all 60 simulated PISNe, and consider the models for the other masses in different simulations.  In order to take into account the possible model dependence of the peak brightness, a Gaussian random variable with 1 mag FWHM is added to the peak magnitude of each random event in the simulation.

In each simulation we use the same ``strong'' and ``weak'' detection criteria as defined by \citet{rv}: ``strong'' means simultaneous detection in all four NIRCam bands, while ``weak'' means detection in at least 2 bands.

Since the random distribution of $\sim 60$ SNe may suffer from biases due to low number statistics, we computed 10 different realization of the 60 randomly distributed PISNe from the same models, and the final redshift distribution of the events that passed the detection criteria (see below) are estimated as the mean values from these 10 different random samples.

Figure~\ref{fig:pisn-simul-strongweak} displays the statistics of the simulations for the zero metallictiy models. ``Strong'' detections are shown in the upper row, where the three subpanels corresponds to the 200, 225 and 250 M$_\odot$ masses, respectively, while the ``weak'' detections are in the lower row. As visible, detection in all four NIRCam filters is not expected beyond $z \sim 4$ according to these PISN models. 2-band detections are less restrictive, so they might occur up to $z \sim 5$, but not beyond that. 

Second, we use the Pop II metallicity PISN model having 250~M$_\odot$ mass by \citet{ch19} in 10 different random realizations of the same simulation as described above (again, assuming for simplicity that all PISNe have the same mass). Because the peak absolute magnitudes of these models ($\sim -20$ AB mag) are $\sim 3$ magnitudes brighter than that of the zero metallicity models, such PISNe could be detected at larger distances, thus, at higher redshifts. Figure~\ref{fig:pisn-simul-distr} displays the number of ``weak'' detections in the four NIRCam bands. As expected, these numbers are somewhat higher than those presented in Figure~\ref{fig:pisn-simul-strongweak}, but the region of detectability does not extend significantly beyond $z\sim 5$. 
Since the ``strong'' detection is mostly constrained by the F150W filter (see Figure~\ref{fig:popii-lc2}), the redshift distribution of the ``strong'' detections would look like the same as the upper left panel of Figure~\ref{fig:pisn-simul-strongweak} for all bands. 
Figure~\ref{fig:pisn-simul-strong} shows the efficiency of the ``strong'' detections. It is seen that for $z < 3$ it is practically 100 \%, i.e. all PISNe are expected to be detectable in all 4 bands. However, it quickly decreases beyond $z \sim 4$, and reach practically zero around $z \sim 5$, meaning that these kind of PISNe at such redshifts could be detected only in the longer wavelength bands. 

\begin{table}
\label{tab:finalnumbers}
\caption{The predicted number of PISN discoveries in the FLARE survey}
\centering
\begin{tabular}{lcc}
\hline
z & $N_{strong}$ & $N_{weak}$ \\
\hline
1 & $6 (\pm 2)$ & $8 (\pm 1)$  \\
2 & $5 (\pm 3)$ & $10 (\pm 3)$  \\
3 & $3 (\pm 3)$ & $6 (\pm 4)$ \\
4 & $1 (\pm 1)$ & $4 (\pm 3)$ \\
5 & $0 (\pm 1)$ & $1 (\pm 1)$ \\ 
\hline
\end{tabular}
\end{table}

Finally, we estimate the likely number of detected PISNe  having different masses and metallicities (assuming 100 \% discovery efficiency) simply by averaging the numbers from the four sets of simulations shown above. Table~\ref{tab:finalnumbers} contains these numbers as a function of redshift, that can be considered as our prediction for the PISN outcome of the {\it JWST} FLARE survey.
(The $z = 1$ row includes PISN in the range $z =$ 0 - 1, and weak detections include the strong detections.)

\section{Discussion}

Even though the numbers in Figure~\ref{fig:pisn-counts}, \ref{fig:pisn-simul-strongweak} and \ref{fig:pisn-simul-distr} suggest that a few PISNe per redshift bin could be discovered with {\em JWST}/NIRCam, there are several caveats that should be kept in mind. 

Our estimates for the redshift distribution of the expected number of PISNe are based our current understanding of the (redshift-dependent) star-formation rate and initial mass function. (cf. Section~\ref{sec:rates2}). At high redshifts these are poorly constrained and may suffer from higher uncertainties. 

Also, we know {\it a priori} neither the mass nor the metallicity of any real PISN, provided they indeed occur at higher redshifts. Therefore, even though we combined several different models for such events, it is uncertain whether the real PISNe would be similar to these models or not.

As it is seen in Figure~\ref{fig:pisn-counts}, there seems to be little chance to detect any Pop III events with {\em JWST}, at least with the conditions of the planned FLARE survey, because their expected numbers do not reach 1 in any redshift bin. On the contrary, Pop II events seem to be more numerous, so the discovery of such PISNe is more probable.

The light curve plots in Figures~1-2, 11-12
suggest that the observer's frame evolution of such events will be slow, on the timescale of several $100$ days, as it is stretched by a factor of $1+z$ due to time dilation. Thus, if such a transient pops up within the FoV, it will likely remain visible during the whole survey. However, the classification of these new events will require ``stong'' detections, i.e. detectability in all four NIRCam filters. The predicted number of such PISNe (Table~\ref{tab:finalnumbers}) is only a few per redshift bin, and these numbers are based on assuming 100 \% discovery efficiency. In reality, it is unlikely that the survey will be able to catch all such events that occur within the FoV. Nevertheless, since the total number of ``strong'' detections is $\sim 15$, even assuming a rather low, 10 percent discovery efficiency would mean that at least 1-2 PISNe might be found with {\em JWST}.

The simulations presented in Section~\ref{sec:simul} highlight that the number of ``strong'' detections is not expected to be significant for events beyond $z\sim 4$ redshift. Because the proposed photometric classification of such objects is based on the color-color diagram (Section~\ref{sec:cc}), this constraint also limits the number of potential discoveries. On the other hand, the proposed 90 days observing cadence of the FLARE survey does not seem to be a strong limitation, as PISNe at $z > 1$ are slowly evolving, as mentioned above. 

Finally, our proposed observational prescription for the discovery of PISNe can be summarized as follows: find a new source within the FoV that is 
\begin{itemize}
    \item{detected in all 4 bands,}
    \item{have the expected brightness and color, according to Figures~\ref{fig:popii-lc2} and \ref{fig:color-color}, and}
    \item{remains detectable with NIRCam for more than 1 year.}
\end{itemize}
Such a transient source could be either a SLSN or a PISN.

\section{Conclusion}

Inspite of their importance to the early Universe, not much is known about the properties of Pop III stars, as not even JWST or 30-40 m class telescopes or the Extremely Large Telescope (ELT) will be able to see them \citep{rz10}. 
Recent discovery of stars above 150 \Ms in the star cluster R136 \citep{r136} and detection of PISN candidates SN 2007bi at z=0.123 \citep{gy09} and SN 2213-1745 at redshift $z=2.05$ \citep{cooke12}
have excited interest in this exotic explosion mechanism and challenged current theories. 
Pair Instability Supernovae, Pulsating PISN will be visible to JWST and ELTs at $z > 20$. Rotating PISN will be visible out to $z \sim 10$.

In this paper we considered Pair Instability Supernovae (PISNe), having both Pop II (sub-solar) and Pop III (near zero) metallicity, and examined whether such events could be detected with the planned FLARE survey with {\em JWST}/NIRCam \citep{flare}. 

Synthetic magnitudes and colors were derived from various PISN models, including the ones published recently by \citet{ch19} as well as new calculations shown in this paper. The PISN rates and their expected numbers within the FLARE field-of-view are inferred in a similar way as in \citet{rv} for SLSNe. 

It is found that PoP II PISNe having $M > 200$ $M_\odot$ seem to be detectable with {\em JWST}/NIRCam down to 27 AB-mag below $z < 4$. Their expected numbers vary between 1 and 10 per unit redshift within $0 < z < 5$ (Figure~\ref{fig:pisn-counts}, \ref{fig:pisn-simul-strongweak} and \ref{fig:pisn-simul-distr}). However, current models of Pop III PISNe predict too faint peak magnitudes to be detectable with {\em JWST}/NIRCam with the proposed observational conditions. Thus, they are more likely to escape detection. The probability of their detection is also decreased by their expected numbers during a 3 year-long survey, which turned out to be below 1 (Figure~\ref{fig:pisn-counts}). 

It is emphasized that all these conclusions are based on state-of-the-art models of our current understanding of PISN explosions. It is also possible that real PISNe, if exist, are much brighter than these models predict. If this were the case, then the detection of even one such event would be a breakthrough in understanding star formation and stellar evolution in the early Universe.

\acknowledgments
The authors express their thanks to Prof. Daniel J. Whalen for providing PopII and PopIII PISN models and for enlightening discussions. We are grateful to the Referee for valuable comments.  
ER and JV were supported by the project "Transient Astrophysical Objects" GINOP 2.3.2-15-2016-00033 of the National Research, Development and Innovation Office (NKFIH), Hungary, funded by the European Union.  

\vspace{5mm}
\facilities{JWST}

\software{GENEVA (Eggenberger et al. 2008), KEPLER (Woosley et al. 2002), FLASH (Fryxell et al. 2000; Dubey et al. 2009), MESA (Paxton et al. 2011, 2013, 2015, 2018, 2019), RAGE (Gittings et al. 2008), SPECTRUM (Frey et al. 2013), sncosmo (Barbary et al. 2016)}

\bibliographystyle{apj}

\bibliography{refs}

\section{Appendix}

Models of Pop III stellar evolution predict that metal-free stars must have initial masses of 140 - 260 \Ms for PISN 
(Heger and Woosley 2002; although Chatzopoulos and Wheeler, 2012 extended the lower mass limit to 85 \Ms if the star rotates).
Here we show additional light curves of the zero metallicity Pop III PISN models in the various NIRCAM bandpasses for 175, 225 and 250 \Ms 
progenitors.

\begin{figure*}[hbt!]
    \centering
    \includegraphics[width=13cm]{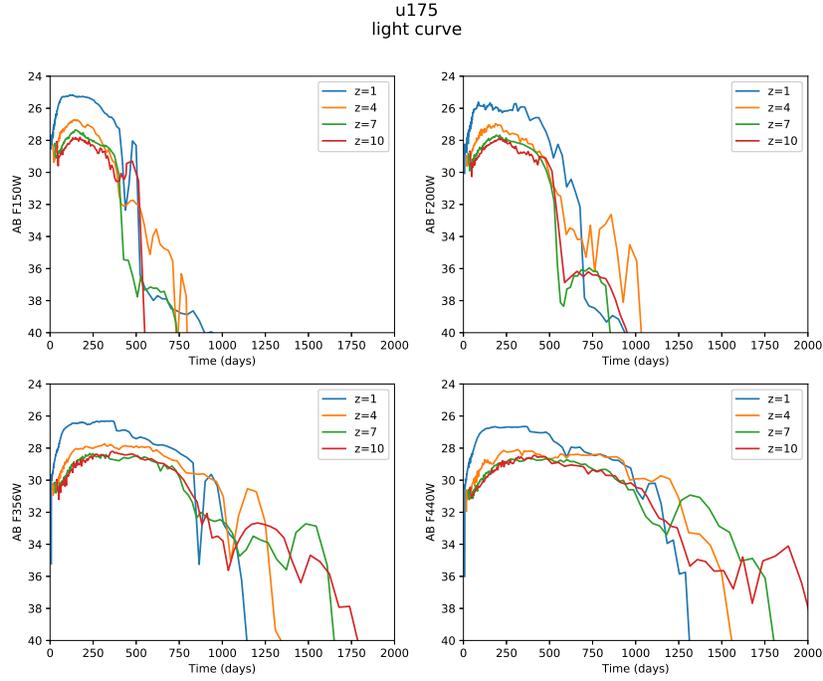}
    \caption{Same as Figure 1 upper panel but for the 175 \Ms model. }
    \label{fig:popiii-a1}
\end{figure*}

\begin{figure*}[hbt!]
    \centering
    \includegraphics[width=13cm]{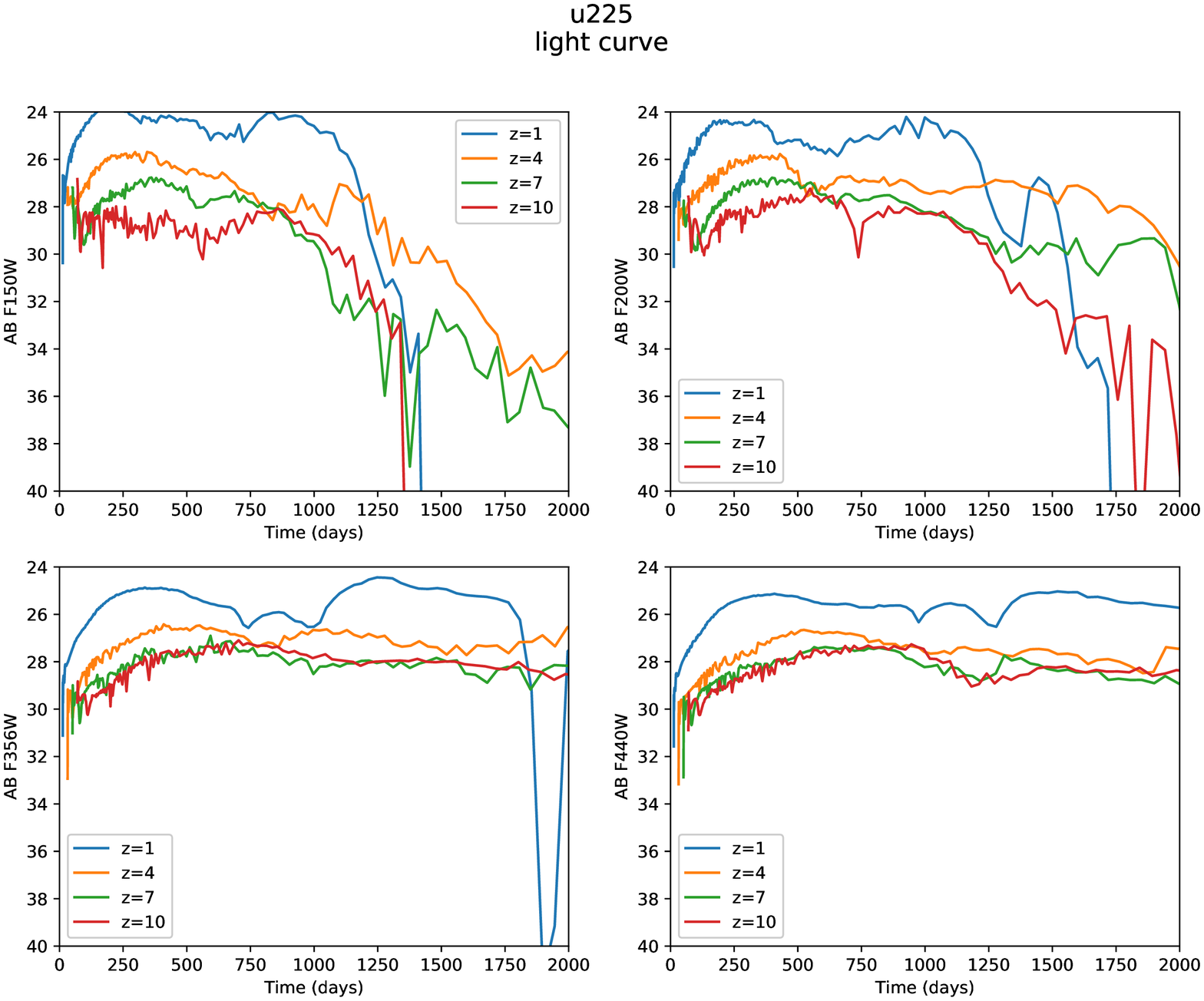}
    \includegraphics[width=13cm]{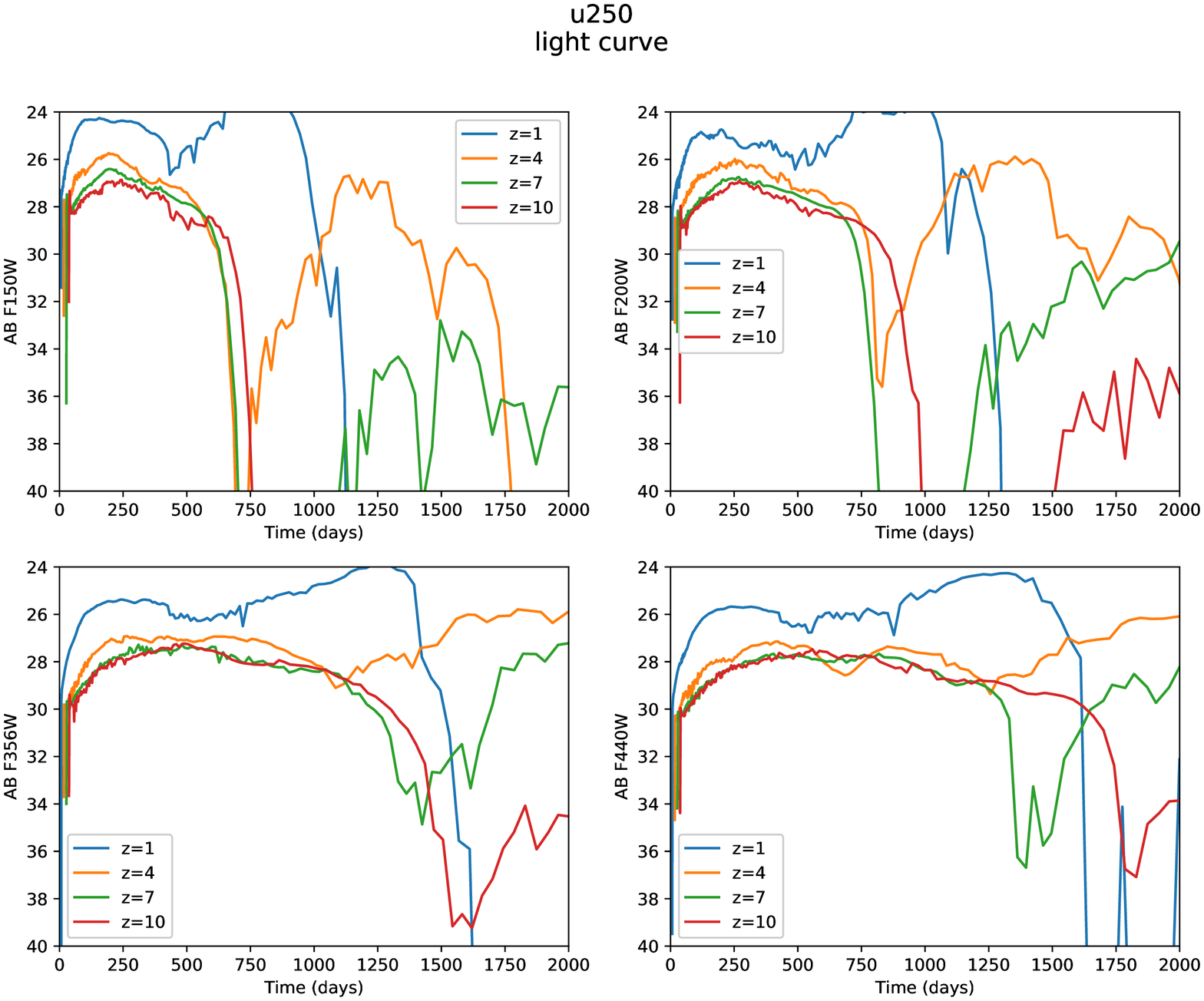}
    \caption{Same as Figure 11 but for the 225 (top panel) and 250 \Ms (bottom panel) models. }
    \label{fig:popii-a2}
\end{figure*}

\end{document}